\documentstyle[12pt]{article}  

\begin{document}

\newtheorem{thm}{Theorem}
\newtheorem{lemma}{Lemma}

 %
 %
 %
 %

\newcommand{\lap}{\bigtriangleup}
\def\be{\begin{equation}}
\def\ee{\end{equation}}
\def\bea{\begin{eqnarray}}
\def\eea{\end{eqnarray}}
\def\beas{\begin{eqnarray*}}
\def\eeas{\end{eqnarray*}}

\def\r{\rho}
\def\s{\sigma}
\def\e{\epsilon}
\def\z{\zeta}
\def\g{\gamma}
\def\x{\xi}
\def\m{\mu}
\def\l{\lambda}

\def\d#1{\partial_{#1}}
\def\dt{\partial_t}
\def\dx{\partial_x}
\def\dv{ \partial_v }
\def\R{{\rm I\kern-.1567em R}}
\def\supp{\mbox{\rm supp}\,} 

\def\open#1{\setbox0=\hbox{$#1$}
\baselineskip = 0pt
\vbox{\hbox{\hspace*{0.5 \wd0}\tiny $\circ$}\hbox{$#1$}} 
\baselineskip = 12pt\!}

\def\n#1{\vert #1 \vert}
\def\nn#1{{\Vert #1 \Vert}}
\def\prf{\noindent
         {\bf Proof:\ }}
\def\prfe{\hspace*{\fill} $\bullet$ 

\smallskip \noindent}

\title{Static shells for the Vlasov-Poisson and Vlasov-Einstein systems} 
\author{Gerhard Rein\\
        Mathematisches Institut der Universit\"at M\"unchen\\
        Theresienstr.\ 39, 80333 M\"unchen, Germany}  
\date{}
\maketitle
\begin{abstract}
We prove the existence of static, spherically symmetric
solutions of the stellar dynamic Vlasov-Poisson and Vlasov-Einstein
systems, which have the property that their spatial support is
a finite, spherically symmetric shell with a vacuum region at the
center. 
\end{abstract}
\section{Introduction}
\setcounter{equation}{0}
Large stellar systems such as galaxies or globular clusters
can be described by a density function  $f\geq 0$ on phase space.
If collisions among the stars are neglected, $f$ satisfies the 
so-called Vlasov or Liouville
equation, which is then coupled to field equations for the gravitational
interaction. Depending on whether one chooses a Newtonian or a general
relativistic setting, the resulting nonlinear system of partial
differential equations is the so-called Vlasov-Poisson
or the Vlasov-Einstein system, respectively. In the present note we 
are interested in time independent spherically symmetric
solutions of these systems. We call these solutions static, since 
due to the spherical symmetry their
average velocity vanishes everywhere.
The Vlasov-Poisson system then takes the following form:
\be \label{nv}
v\cdot \nabla_x f - \nabla_x U \cdot \nabla_v f =0,
\ee
\be \label{nf} 
\frac{1}{r^2}(r^2 U')'= 4\pi \r
\ee
where
\be \label{nrdef}
\r (r) = \r (x)= \int f(x,v)\,dv .
\ee
Here $x, v \in \R^3$ denote position and momentum, $r=\n{x}$,
$'$ denotes the derivative with respect to $r$,
$f=f(x,v)$ must be spherically symmetric, i.~e., $f(x,v)=f(Ax,Av)$
for every rotation $A \in \mbox{SO}(3)$, 
$\r(x)=\r(r)$ denotes the spatial mass density of the ensemble
and $U(x)=U(r)$ is the induced gravitational potential.
We assume that all particles in the ensemble have the same mass
which---like all other physical constants---is set to unity.

Under the corresponding assumptions the Vlasov-Einstein system
takes the form
\be \label{rv}
\frac{v}{\sqrt{1+ v^2}}\cdot \dx f -
\sqrt{1+ v^2}\, \m'\, \frac{x}{r} \cdot \dv f =0,
\ee
\bea
e^{-2\l} (2 r \l' -1) +1 &=& 8\pi r^2 \r ,\label{rf1}\\
e^{-2\l} (2 r \m' +1) -1 &=& 8\pi r^2 p,  \label{rf2}
\eea
where
\bea
\r(r) = \r(x) 
&=& 
\int \sqrt{1+ v^2} f(x,v)\,dv ,
\label{rrdef}\\
p(r) =p(x)  
&=& 
\int\left(\frac{x\cdot v}{r}\right)^2 f(x,v)
\frac{dv}{\sqrt{1+v^2}} 
\label{rpdef}
\eea
denote the spatial density of mass-energy and radial pressure, respectively.
If
$x=r(\sin \theta \cos\phi, \sin \theta \sin \phi, \cos\theta)$ then
the spacetime metric is given by
\[
ds^2 = - e^{2\m} dt^2 + e^{2\l} dr^2 + r^2( d\theta^2 + \sin^2\theta d\phi^2).
\]
As boundary conditions we require asymptotic flatness, i.~e.,
\be \label{rbc1}
\lim_{r\to\infty} \l(r) = \lim_{r\to\infty}\m(r) =0 ,
\ee
and a regular center, i.~e.,
\be \label{rbc2}
\l(0)=0 .
\ee
For the Vlasov-Poisson system the corresponding boundary condition
is
\be \label{nbc}
\lim_{r\to \infty} U(r) =0.
\ee
All solutions of the above systems known so far have the property
that the support of $\r$ contains a ball about the center;
the only steady states where the support does not equal such a
ball are the axially symmetric ones obtained in \cite{R2}.
The purpose
of the present note is to construct solutions whose support is a 
finite, spherically symmetric shell so that they have a 
vacuum region at the center. Given the fact that the dynamical
behaviour of both the Vlasov-Poisson and the Vlasov-Einstein systems
is far from being understood, static solutions with new structural
properties are of interest in themselves. However, there is also
a more specific motivation for the present investigation:
In \cite{RRS2} the gravitational collapse of spherically symmetric
solutions of the Vlasov-Einstein system is investigated numerically.
Static solutions provide useful test cases for the corresponding numerical
scheme. Since the center of symmetry is particularly difficult to handle,
it is important to have static solutions both with matter and with
vacuum at the center in order to assess the performance of the numerical
scheme. As shown in \cite{RRS1} symmetric solutions of the time dependent
problem which vanish near the center remain smooth. Thus, staying away 
from the center avoids analytic as well as numeric difficulties.

The way to construct such steady states is now described. 
Since the system is time independent, the particle energy
must be a conserved quantity, since it is spherically symmetric, 
the same is true for the modulus of angular momentum.
Indeed, the quantities
\be \label{nconq}
E = E(x,v) = \frac{1}{2} v^2 + U(x),\ L = L(x,v)=\n{x\times v}^2
\ee
are constant along solutions of the characteristic equations
\[
\dot x = v,\ \dot v = -\nabla U(x)
\]
of the nonrelativistic Vlasov equation (\ref{nv}), and
\be \label{rconq}
E = E(x,v) = e^{\m (r)} \sqrt{1+v^2},\ L = L(x,v)=\n{x \times v}^2
\ee
are constant along characteristics of the relativistic
Vlasov equation (\ref{rv}). 
Therefore, the ansatz 
\be \label{ansatz}
f(x,v) = \Phi (E,L)
\ee
satisfies the corresponding Vlasov equation and reduces
the system to the field equation(s),
where the source terms $\r$ or $\r$ and $p$ now become functionals of
$U$ or $\mu$, which are obtained by substituting the ansatz
(\ref{ansatz}) into (\ref{nrdef}) or (\ref{rrdef}), (\ref{rpdef})
respectively.
In passing we note that
every static, spherically symmetric solution 
of the Vlasov-Poisson system must be of the form
(\ref{ansatz}), cf.\ \cite{BFH}. 
For the Vlasov-Einstein system this result, usually
referred to as Jeans' Theorem, is not established.

One can easily see that $\r$ becomes a decreasing function of $r$
if $f$ is a function of the particle energy $E$ only, the so-called
isotropic case. Thus, to obtain a nontrivial solution with a vacuum
region at the center, $f$ must also depend on the angular momentum $L$,
and it must vanish for $L$ small, say for $L \leq L_0$ for some
$L_0>0$. Once a solution of the field equation(s)
is obtained, which has 
a vacuum region at the center, the main difficulty is to show that
the support of the solution is actually bounded and the solution
leads to a model with finite mass
\[
M = \int \r(x)\, dx < \infty;
\]
in the case of the Vlasov-Einstein system this quantity is the so-called
ADM mass.
Finiteness of mass and support are obtained as follows. We take an 
ansatz function $\Phi$, depending on the parameter $L_0$ in such a way
that for $L_0=0$ known results give the existence of a solution
with finite mass and finite support, in this case a ball about
the center. Then a perturbation argument in $L_0$ is used to show that these
properties persist also for $L_0>0$ but small. 
The smallness assumption on $L_0$ can then be removed by a scaling
argument.
The details of this procedure together with the precise statements
of our results are given in the next section for the Vlasov-Poisson
system, and in the last section for the Vlasov-Einstein system.

Before we go into this, we give a brief overview of the literature
on the Vlasov-Poisson and the Vlasov-Einstein systems, starting
with the former. We restrict ourselves to the stellar dynamics
case; the plasma physics case, where the sign
in the Poisson equation is reversed, is omitted.
Global existence of classical solutions has been established in 
\cite{Pf}, cf.\ also \cite{LP,Sch}.
As far as the existence of stationary solutions of
the Vlasov-Poisson system is concerned we mention \cite{BFH,BP,R2}.
The main result on the initial value problem
for the Vlasov-Einstein system is a global 
existence theorem for small, spherically symmetric data \cite{RR1}.
Spherically symmetric steady states for the Vlasov-Einstein system
are constructed in \cite{R1,RR2}.

\section{The nonrelativistic case}
\setcounter{equation}{0}
Throughout this section we fix two parameters $k, l \in \R$ with
\[
k>-1,\ l>-1,\ k+l+\frac{1}{2} \geq 0,\ k< 3l+\frac{7}{2}.
\]
We make the ansatz
\be \label{nansatz}
f(x,v) = c_0\, (E_0 - E)_+^k (L-L_0)_+^l
\ee
where $E$ and $L$ are defined as in (\ref{nconq}), $(\cdot)_+$ denotes the
positive part of the argument,
and  $c_0 >0,\ E_0 < 0,\ L_0 \geq 0$ .
It is a straight forward computation to show that with this ansatz
\be \label{nrrdef}
\r(r)= r^{2l} g\left(U(r)+\frac{L_0}{2r^2}\right),
\ee
where
\[
g(u):=c_0 c_{kl}  
(E_0 - u)_+^{k+l+\frac{3}{2}}
\]
and
\[
c_{kl}:= 2^{l+\frac{3}{2}} \pi \int_0^1 \frac{s^l}{\sqrt{1-s}}ds
\int_0^1 s^{l+\frac{1}{2}} (1-s)^k ds ,
\]
and we have to solve
\be \label{nrf}
\frac{1}{r^2} (r^2 U')' = 4 \pi r^{2l}
g\left(U+\frac{L_0}{2r^2}\right),\ r >0.
\ee 
The exponents $k$ and $l$ are kept fixed, while the parameters
$c_0$, $E_0$, and $L_0$ may vary during our argument. The following
theorem is the main result of the present section:

\begin{thm}
Let $M>0$ and $R_0>0$. Then there exists a static,
spherically symmetric solution $(f,\r,U)$
of the Vlasov-Poisson system (\ref{nv}), (\ref{nf}), (\ref{nrdef}),
where $f$ and $\r$ depend on $U$ via (\ref{nansatz}) and (\ref{nrrdef}).
$U\in C^2([0,\infty[)\cap C^2(\R^3)$ is a solution of (\ref{nrf}) 
satisfying the boundary condition (\ref{nbc}),
$\r \in C^1([0,\infty[)\cap C^1(\R^3)$
has total mass $M$, and
$\supp \r = [R_i,R_0]$ for some $R_i \in [0,R_0[$,
where $R_i>0$ provided $L_0>0$.
Instead of prescribing $M>0$ and $R_0>0$ one
may also prescribe $M>0$ and $R_i > 0$.
If $L_0=0$ and $0 \neq l \leq 1/2$ then the asserted regularity holds only
on $\R^3\setminus\{0\}$.
\end{thm}
Note that we identify spherically symmetric
functions of $x$ with the corresponding functions
of $r=\n{x}$. 

\noindent
{\em Proof:}
Let us fix some $E_0$ and $c_0=1$, and consider $L_0=0$ first.
Then for $U_0(0) < E_0$ prescribed there exists a unique solution
$U_0$ of
\[
\frac{1}{r^2}(r^2 U')'=4\, \pi\, r^{2l} g(U),
\]
cf.\ \cite{BFH}; this solution need not satisfy
the boundary condition (\ref{nbc}), but we will take care of that later.
By \cite{BFH}, $U_0$ induces a steady state with finite mass and
finite support, which means that for some $R_0>0$ we have 
$U_0(r) > E_0$ for all $r\geq R_0$.
For $L_0>0$ we define 
\[
U_{L_0}(r)=U_0(0),\ 0\leq r \leq r_{L_0} 
\]
where
\[
r_{L_0}:= \sqrt{\frac{L_0}{2(E_0-U_0(0))}};
\]
note that $U_{L_0}(r)+\frac{L_0}{2r^2} > E_0$ on $]0,r_{L_0}[$,
so the right hand side of the Poisson equation vanishes on that
interval. Now extend this towards the right by the solution 
of (\ref{nrf}) with
$U_{L_0}(r_{L_0})=U_0(0),\ U'_{L_0}(r_{L_0})=0$. 
The latter exists on $[0,\infty[$,
which can be shown using \cite{BFH}, but this also follows from the 
arguments below. Upon integrating the Poisson equation we find
\beas
U'_0(r)
&=&
\frac{4 \pi}{r^2} \int_0^r s^{2+2l}g(U_0(s))\,ds,\\
U'_{L_0}(r)
&=&
\frac{4 \pi}{r^2} \int_0^r s^{2+2l}
g\left(U_{L_0}(s)+\frac{L_0}{2 s^2}\right)\,ds,
\eeas
where the latter integral is zero for $r\leq r_{L_0}$. 
Now observe that $U_0(r) \geq  U_0(0)$ and $U_{L_0}(r)\geq U_0(0) $ 
for $r \geq 0$,
and on the set $[U_0(0),\infty[$ the function $g$ is bounded and Lipschitz;
recall that we assume $k+l+1/2 \geq 0$.
For $0\leq r\leq r_{L_0}$ we have
\[
\left| U'_0(r) - U'_{L_0}(r)\right|
\leq
\frac{C}{r^2}\int_0^r s^{2+2l}ds = C r^{2l+1}
\]
and for $r\geq r_{L_0}$
\beas
\left| U'_0(r) - U'_{L_0}(r)\right|
&\leq& 
U'_0(r_{L_0})
 + \frac{C}{r^2} \int_{r_{L_0}}^r  s^{2+2l}
\left(\left|U_0(s)-U_{L_0}(s)\right| +\frac{L_0}{2s^2}\right)\,ds\\
&\leq&
C r_{L_0}^{2l+1} +
\frac{C}{r^2}L_0 r_{L_0}^{2\e-2}\int_0^r s^{2+2l-2\e}ds\\
&&
{}
+\frac{C}{r^2}\int_{r_{L_0}}^r s^{2+2l}\left|U_0(s)-U_{L_0}(s)\right|\, ds\\
&\leq&
C L_0^{\e} r^{2l+1-2\e} +
\frac{C}{r^2}\int_{r_{L_0}}^r s^{2+2l}\left|U_0(s)-U_{L_0}(s)\right|\, ds,
\eeas
where $\epsilon>0$ is such that $1+l-\e >0$;
constants denoted by $C$ may depend on $U_0$ and $R_0$, 
but not on $r$ or $L_0$,
and may change from line to line. 
Thus
\beas
\left| U_0(r) - U_{L_0}(r) \right| 
&\leq&
C L_0^\e + C\,
\int_{r_{L_0}}^r\frac{1}{s^2}
\int_{r_{L_0}}^s \s^{2+2l}
\left|U_0(\s)-U_{L_0}(\s)\right|\, d\s\,ds \\
&\leq&
C L_0^\e + C\, \int_0^r s^{2l+1} \sup_{0\leq\s\leq s}
\left|U_0(\s)-U_{L_0}(\s)\right|\,ds ,
\eeas
and the latter inequality holds
for $0\leq r\leq R_0$. By Gronwall's Lemma,
\[
\left| U_0(r) - U_{L_0}(r) \right| \leq C L_0^\e,\ 0 \leq r \leq R_0;
\]
note that
$s^{2l+1}$ is integrable over this interval. In particular, 
$U_{L_0}$ must exist at least on this interval  
for $L_0$ small. We conclude that 
\[
U_{L_0}(R_0) > E_0
\]
for $L_0>0$ sufficiently small, and by monotonicity,
\[
U_{L_0}(r) + \frac{L_0}{2r^2} > E_0,\ r \geq R_0.
\]
We fix a small $L_0>0$ and let $R_i=r_{L_0}$ and $U=U_{L_0}$ etc.
Then $U$ leads to a steady state with finite radius and finite mass.
The regularity assertions are obvious in the case $L_0>0$. In the case
$L_0=0$ the exponent $l$ has to be restricted in such a way that
$\r'(0)=U'(0)=0$ and $U''(0)$ exists.

Let $V(r) = U(r) + \frac{L_0}{2r^2}$.
Then $V(R_i)=E_0$ and $V'(R_i) <0$, whence
$V(r) < E_0$ in a right neighborhood of $R_i$.
Thus the induced mass density is nontrivial
and $R_i$ is the radius of the inner boundary of its support.
If we define $M(r)= 4 \pi \int_0^r s^2 \r (s)\, ds$
then $M(r)$ is increasing with $M(r) >0$ for $r>R_i$, and
\[
V'(r) = \frac{M(r)}{r^2} - \frac{L_0}{r^3}.
\]
Thus there can be at most one value of $r>R_i$ where $V'(r)$ changes
sign, and if we define $R_0=\inf\{r>R_i | V(r) = E_0\}$, this set being 
nonempty by what we showed above,
then $R_0 > R_i$, and $\supp \r = [R_i,R_0]$, i.~e., the support of
the steady state consists of a single shell and not several nested
ones.

So far the boundary condition (\ref{nbc}) need not be satisfied,
but $U(\infty)= \lim_{r \to \infty} U(r) > E_0$ exists,
and by slightly abusing notation we can redefine
\[
U = U - U(\infty),\
E_0 = E_0 - U(\infty).
\]
This leaves the distribution function $f$ unchanged, and in addition
(\ref{nbc}) is now satisfied.

Finally we note that if
$f$ is a solution of the static Vlasov-Poisson system with mass
$M$ and $\supp \r = [R_i,R_0]$ then the function
\[
f_{\l,\m}(x,v) = \g^3 \l^{-1} f(\g x, \g \l^{-1}v)
\]
is also a solution, with mass
$M(\l,\g) = \l^2 \g^{-3} M$
and support of the spatial density equal to 
$[R_i/\g, R_0/\g]$. Choosing $\gamma$ and $\lambda$ appropriately
any prescribed value for $R_0$ (or $R_i$) and $M$ can be obtained.
Obviously, the constants $c_0,\ E_0$, and $L_0$ in the original ansatz
(\ref{nansatz}) are changed by this scaling, but the boundary condition 
(\ref{nbc}) is not. 

\section{The relativistic case}
\setcounter{equation}{0}
Throughout this section we fix two parameters $k, l \in \R$ with
\[
k\geq 0,\ l>-\frac{1}{2},\ k< 3l+\frac{7}{2}.
\]
We again make the ansatz
\be \label{ransatz}
f(x,v) = c_0 (E_0-E)_+^k (L-L_0)_+^l
\ee
where $E$ and $L$ are now defined as in (\ref{rconq}),
and $c_0,\ E_0 >0$, $L_0 \geq 0$.
With this ansatz
\bea 
\r(r)
&=&
r^{2l} e^{-(2l+4)\mu} g\left(e^\mu\sqrt{1+L_0/r^2}\right),
\label{rrrdef}\\
p(r)
&=&
r^{2l} e^{-(2l+4)\mu} 
h\left(e^\mu\sqrt{1+L_0/r^2}\right),
\label{prrdef}
\eea
where 
\bea
g(u)
&:=&
c_0 c_l
\int_u^\infty (E_0-E)_+^k E^2 (E^2-u^2)^{l+1/2} dE, \label{rgdef}\\
h(u)
&:=&
\frac{c_0 c_l}{2l+3}
\int_u^\infty (E_0-E)_+^k (E^2-u^2)^{l+3/2} dE, \label{rhdef}
\eea
and
\[
c_l:=2 \pi \int_0^1 \frac{s^l}{\sqrt{1-s}}ds .
\]
Taking into account the boundary condition (\ref{rbc2}) we can integrate
the field equation (\ref{rf1}) to obtain
\[
e^{-2\lambda} = 1 - \frac{8 \pi}{r}\int_0^rs^2 \r (s)\, ds,
\]
and substituting this into (\ref{rf2}) reduces the static, spherically
symmetric Vlasov-Einstein system to the equation
\be 
\mu'(r) = \left(1-\frac{8 \pi}{r}\int_0^rs^2 \r (s)\, ds\right)^{-1}
\left(4\pi\,r\,p(r) + \frac{4\pi}{r^2}\int_0^rs^2 \r (s)\, ds\right),
\label{rrf}
\ee
where $\r$ and $p$ are now functionals of $\mu$ given by 
(\ref{rrrdef}), (\ref{prrdef}), (\ref{rgdef}), (\ref{rhdef}).
The following theorem is the main result of the present section:

\begin{thm}
There exists a static,
spherically symmetric solution $(f,\r,p,\l,\m)$
of the Vlasov-Einstein system (\ref{rv}), (\ref{rf1}), (\ref{rf2}),
(\ref{rrdef}), (\ref{rpdef}),
where $f,\ \r$, and $p$ depend on $\mu$ via (\ref{ransatz}),
(\ref{rrrdef}), and (\ref{prrdef}) in a neighborhood of their support.
$\lambda,\mu\in C^2([0,\infty[)\cap C^2(\R^3)$ 
satisfy the boundary conditions (\ref{rbc1}), (\ref{rbc2}), and $\mu$
is a solution of (\ref{rrf}).
$\r,p \in C^1([0,\infty[)\cap C^1(\R^3)$ with
$\supp \r =\supp p = [R_i,R_0]$ for some $0 \leq R_i < R_0 < \infty$,
where $R_i>0$ provided $L_0>0$.
The ADM mass $M$
is finite, and one can prescribe $M>0$ or $R_0>0$ or $R_i>0$.
If $L_0=0$ and $0 \neq l \leq 1/2$ then the asserted regularity holds only
on $\R^3\setminus\{0\}$.
\end{thm}

\noindent
{\em Proof}: Consider first the case
$L_0=0$. As was shown in \cite{R1} there exists
$E_0 >0$ and a solution $\mu_0$ of (\ref{rrf}) with 
$e^{\mu_0(0)} < E_0$ and
$e^{\mu_0(R_0)} > E_0$ for some $R_0>0$. We choose $R_0$ such that
$E_0 < e^{\mu_0(R_0)} < E_0 +1$. For any $L_0>0$
we define
\[
r_{L_0} := \sqrt{\frac{L_0}{E_0^2 e^{-2\mu_0(0)} -1}}.
\]
Then
\[
e^{\mu_0(0)}\sqrt{1+L_0/r_{L_0}^2} = E_0,\
e^{\mu_0(0)}\sqrt{1+L_0/r^2} >  E_0,\ r \in [0,r_{L_0}[,
\]
which means that $\mu_{L_0}(r)=\mu_0(0)$ solves (\ref{rrf})
on $[0,r_{L_0}]$ with 
$\rho_{L_0} (r)=p_{L_0} (r) = 0$; in what follows
$\rho_{L_0}$ and $p_{L_0}$ are always given in terms of $\mu_{L_0}$
by (\ref{rrrdef}) and (\ref{prrdef}) respectively.  
By \cite[Thm.~3.1]{R1}  $\mu_{L_0}$ can be extended as a solution
of (\ref{rrf}) for $r\geq r_{L_0}$.
We want to show that $\exp(\mu_{L_0}(R_0)) > E_0$ for $L_0>0$ small
so we may assume that
$\mu_{L_0}(R_0) < \mu_0(R_0)+1$ since otherwise we are done. By 
monotonicity,
\[
\mu_0(0) \leq \mu_0(r),\ \mu_{L_0}(r) < \mu_0(R_0)+1,\ r \in [0,R_0].
\] 
The functions $g$ and $h$ can be shown to be continuously differentiable,
cf.\ \cite[Lemma 2.1]{RR2}, and they vanish for $u > E_0$. Thus
\bea
\n{\rho_{L_0} (r) - \r_0(r)}
&\leq&
C r^{2l} \left|e^{-(2l+4) \mu_{L_0}} - e^{-(2l+4) \mu_0}\right|
g\left(e^{\mu_{L_0}}\sqrt{1+L_0/r^2}\right)
\nonumber \\
&&
{}+ C r^{2l} e^{-(2l+4)\mu_0}
\left|g\left(e^{\mu_{L_0}}\sqrt{1+L_0/r^2}\right)
- g\left(e^{\mu_0}\right)\right|
\nonumber \\
&\leq&
C \left(r^{2l} \n{\mu_{L_0}(r)-\mu_0(r)} + r^{2l-1} \sqrt{L_0}\right),\
r\in ]0,R_0], \label{rhodiff}
\eea
and similarly
\be \label{pdiff}
\n{p_{L_0} (r) - p_0(r)} 
\leq
C \left(r^{2l} \n{\mu_{L_0}(r)-\mu_0(r)} + r^{2l-1} \sqrt{L_0}\right),\
r\in ]0,R_0].
\ee
Constants denoted by $C$ may depend on $\mu_0$ and $R_0$, but never on
$r$ or $L_0$, and may change from line to line.
From (\ref{rrf}) we obtain the estimate
\beas
\n{\mu'_{L_0} (r) - \mu'_0(r)} 
&\leq& 
\left|\left(1-\frac{8 \pi}{r}\int_0^rs^2 \r_{L_0} (s)\, ds\right)^{-1}
- \left(1-\frac{8 \pi}{r}\int_0^rs^2 \r_0 (s)\, ds\right)^{-1}\right|\\
&& \hspace{125pt}
\left(4\pi\,r\,p_{L_0}(r) + 
\frac{4\pi}{r^2}\int_0^rs^2 \r_{L_0} (s)\, ds\right),\\
&&
{}+
4\pi\, \left(1-\frac{8 \pi}{r}\int_0^rs^2 \r_0 (s)\, ds\right)^{-1}\\
&& \hspace{20pt}
\left|r\,p_{L_0}(r) + 
\frac{1}{r^2}\int_0^rs^2 \r_{L_0} (s)\, ds
- 
r\,p_0(r) - 
\frac{1}{r^2}\int_0^rs^2 \r_0 (s)\, ds \right|\\
&=&
I + II.
\eeas
Suppose that
\be \label{apest}
\sup_{0\leq r\leq R} \n{\mu_{L_0}(r) - \mu_0(r)} \leq \gamma
\ee
for some $\gamma >0$ and $R\in ]0,R_0]$. Then by (\ref{rhodiff}),
\[
\frac{8\pi}{r}\int_0^r s^2 \n{\r_{L_0}(s)-\r_0(s)}\,ds
\leq 
\frac{C}{r}\int_0^r 
\left(s^{2l+2}\gamma +s^{2l+1}\sqrt{L_0}\right)\,ds
\leq
C_1 \left(\gamma + \sqrt{L_0}\right), 
\]
recall that $l> -1/2$. 
Now we choose $\gamma >0$ such that
\[
\sup_{0<r\leq R_0}\frac{8 \pi}{r}\int_0^r s^2 \r_0 (s)\,ds 
+ 2 C_1 \gamma < 1.
\]
For $L_0\leq \gamma^2$ this implies that
\beas
\sup_{0<r\leq R}\frac{8 \pi}{r}\int_0^r s^2 \r_{L_0} (s)\,ds
&\leq&
\sup_{0<r\leq R}\frac{8 \pi}{r}\int_0^r s^2 \r_0 (s)\,ds
+  2 C_1 \gamma < 1.
\eeas
Thus
\[
\left(1-\frac{8 \pi}{r}\int_0^rs^2 \r_{L_0} (s)\, ds\right)^{-1}
< C,\ 0 <  r \leq R,
\]
provided (\ref{apest}) holds. This allows us to estimate the term $I$ above:
\beas
I
&\leq&
C r^{2l}\int_0^r s^2 \n{\r_{L_0}(s)-\r_0(s)}\,ds \\
&\leq&
C r^{4l+2} \sqrt{L_0} + 
C r^{2l} \int_0^r s^{2l+2}\n{\mu_{L_0}(s)-\mu_0(s)}\,ds.
\eeas
As to the second term we find that
\beas
II
&\leq&
C r \n{p_{L_0}(r)-p_0(r)}
+ \frac{C}{r^2} \int_0^r s^2 
\n{\r_{L_0}(s)-\r_0(s)}\,ds\\
&\leq&
C \sqrt{L_0} r^{2l} +  
C r^{2l+1} \n{\mu_{L_0}(r)-\mu_0(r)} + 
C \int_0^r s^{2l}\n{\mu_{L_0}(s)-\mu_0(s)}\,ds .
\eeas
Thus
\[
\n{\mu_{L_0}(r)-\mu_0(r)}
\leq C \left( \sqrt{L_0} + \int_0^r s^{2l} 
\n{\mu_{L_0}(s)-\mu_0(s)}\,ds\right)
\]
on $[0,R]$, and by Gronwall's lemma,
\[
\n{\mu_{L_0}(r)-\mu_0(r)} \leq C \sqrt{L_0},\ r \in [0,R].
\]
By choosing $L_0$ small we can make sure that (\ref{apest}) holds on $[0,R_0]$
so that the previous estimate holds on $[0,R_0]$ as well. In particular,
$\exp(\mu_{L_0}(R_0)) > E_0$ provided $L_0$ is sufficiently small.
We fix a sufficiently small $L_0$ and let $R_i=r_{L_0}$ and $\mu=\mu_{L_0}$
etc.

It is easy to see that $e^{\mu(r)} \sqrt{1+L_0/r^2} < E_0$
and thus $\r (r) >0$ 
on some interval $]R_i,R[$. Take $R_0 > R_i$
the smallest such $R$ with the property that
$\r=0$ in a right neighbourhood of
$R_0$. It is not clear that 
$e^{\mu(r)} \sqrt{1+L_0/r^2} > E_0$ for all $r>R_0$,
but we can simply extend the solution towards the right of $R_0$
by the corresponding vacuum solution.
Thus, while $f$, $\rho$, and $p$ depend on $\mu$ via
(\ref{ransatz}), (\ref{rrrdef}), and (\ref{prrdef}) 
in a neighborhood of their support, this need not be true for all values 
of $r$.
Clearly, $\mu(\infty)=\lim_{r\to \infty} \mu(r)$ exists. If we redefine
\[
\mu (r) = \mu (r) - \mu(\infty),\ E_0 = E_0 e^{-\mu(\infty)}
\]
we satisfy the boundary condition at infinity while the constant $c_0$
in (\ref{ransatz}) is multiplied by $e^{k \mu(\infty)}$.

Finally, if $f(x,v)$ defines a steady state, so does
\[
f_a (x,v) = a^2 f(ax,v)
\]
for any $a>0$. The rescaled function $f_a$ has spatial support 
$[a^{-1}R_i, a^{-1} R_0]$ and ADM mass
\[
a^{-1} \int f(x,v) \sqrt{1+v^2}\,dv\,dx
\]
which shows that by rescaling a given solution we can get any prescribed
value for the ADM mass, or the inner, or the outer radius.

\noindent
{\bf Final remark:} It would be desirable to have a complete parametrization
of all steady states as constructed in Thms.~1 and 2 (for $k$ and $l$
fixed), that is to say, a result of the form: For every 
$0 \leq R_i < R_0 < \infty$ and $M>0$ there exists a unique steady state
of the form (\ref{nansatz}) or (\ref{ransatz}) with support
$[R_i,R_0]$ and mass $M$. Such a result can be obtained in the 
Vlasov-Poisson case for $L_0=0$, cf.\ \cite{R3}.

\noindent
{\bf Acknowledgement:} I would like to thank the Department
of Mathematics, Indiana University, Bloomington, for its hospitality
during the academic year 1997/98.

\end{document}